\documentclass[aps,prd,preprint,groupedaddress,superscriptaddress,nofootinbib,11pt]{revtex4-1}

\usepackage{amsfonts}
\usepackage{amsmath}
\usepackage{graphicx}
\usepackage{hyperref}
\usepackage{anyfontsize}
\usepackage{ulem}
\usepackage[usenames]{color}
\usepackage{lineno}

\hypersetup{colorlinks, linkcolor = [rgb]{0,0.0,0.5}, citecolor = [rgb]{0,0.0,0.5}, urlcolor = [rgb]{0,0.0,0.5}}

\def\be{\begin{equation}}
\def\ee{\end{equation}}
\def\bea{\begin{eqnarray}}
\def\eea{\end{eqnarray}}
\def\bec{\begin{center}}
\def\enc{\end{center}}
\def\lb{\label}

\newcommand{\req}[1]{(\ref{#1})}
\newcommand{\half}{\textstyle \frac{1}{2}}

\begin{document}

\preprint{SLAC-PUB-17374}

\title{The Isoscalar Mesons and Exotic States in Light Front Holographic QCD}

\newcommand*{\LANZHOU}{Institute of Modern Physics, Chinese Academy of Sciences, Lanzhou 730000, China}\affiliation{\LANZHOU}
\newcommand*{\HEIDELBERG}{Institut f\"ur Theoretische Physik der Universit\"at, D-69120 Heidelberg, Germany}\affiliation{\HEIDELBERG}
\newcommand*{\COSTARICA}{Laboratorio de F\'isica Te\'orica y Computacional, Universidad de Costa Rica, 11501 San Jos\'e, Costa Rica}\affiliation{\COSTARICA}
\newcommand*{\SLAC}{SLAC National Accelerator Laboratory, Stanford University, Stanford, CA 94309, USA}\affiliation{\SLAC}

\author{Liping~Zou}\email{zoulp@impcas.ac.cn}\affiliation{\LANZHOU}
\author{Hans~G\"unter~Dosch}\email{h.g.dosch@thphys.uni-heidelberg.de}\affiliation{\LANZHOU}\affiliation{\HEIDELBERG}
\author{Guy~F.~de~T\'eramond}\email{gdt@asterix.crnet.cr}\affiliation{\COSTARICA}
\author{Stanley~J.~Brodsky}\email{sjbth@slac.stanford.edu}\affiliation{\SLAC}


\begin{abstract}

In this article a systematic quantitative analysis of the isoscalar bosonic states is performed in the framework of supersymmetric light front holographic QCD. It is shown that the spectroscopy of the $\eta$ and $h$ mesons can be well described if  one additional  mass parameter --  which corresponds to  the hard breaking of  chiral $U(1)$  symmetry in standard QCD --   is introduced. The mass difference of the $\eta$ and $\eta'$ isoscalar  mesons is then determined by the strange quark mass content  of the $\eta'$. The theory also predicts the existence of  isoscalar tetraquarks  which are bound states of diquarks and anti-diquarks. The candidates for these exotic isoscalar tetraquarks are identified.  In particular, the $f_0(1500)$ is identified as  isoscalar tetraquark;  the predicted mass value 1.52 GeV agrees with the measured experimental value within the model uncertainties.

\end{abstract}


\maketitle

\section{Introduction}

In a series of recent papers~\cite{deTeramond:2014asa,Dosch:2015nwa,Brodsky:2016yod,Dosch:2015bca,Dosch:2016zdv,Nielsen:2018uyn,Nielsen:2018ytt} it was shown that Light-Front Holographic QCD (LFHQCD)\cite{Brodsky:2003px,deTeramond:2008ht,Brodsky:2014yha}, especially after implementation of superconformal algebra~\cite{deTeramond:2014asa,Dosch:2015nwa} (supersymmetric LFHQCD) explains many spectroscopic and dynamical features of the observed hadrons, thus providing nontrivial analytic connections between the spectroscopy of hadrons  and  the dynamics underlying observables such as their form factors and quark distributions~\cite{deTeramond:2018ecg, Sufian:2018cpj}.  The LFHQCD formalism is based on Lorentz-invariant bound-state light-front Schr\"odinger equations with a single mass parameter appearing in a color-confining potential. Mesons, baryons and tetraquarks are unified as components of a supersymmetric multiplet.

LFHQCD is inspired by the Maldacena conjecture~\cite{Maldacena:1997re}: a  weakly coupled classical 5-dimensional gravitational theory with anti-de Sitter (AdS) metric is a holographical dual to a strongly coupled 4-dimensional quantum gauge field theory defined at the space-time boundary of AdS$_5$. This resulting 4-dimensional field theory  is a superconformal gauge theory in the limit of $N_C \to \infty$ colors. A crucial feature of LFHQCD is the correspondence between the fields  of the 5-dimensional theory and those of the 4-dimensional theory at fixed light-front time $\tau=x^+ = x^0+x^3.$  This remarkable correspondence is based on the observation that the classical equations of motion derived from the action of the 5-dimensional theory, have the identical form as bound-state equations for two massless constituents in light front (LF) quantization. This holographic correspondence is also realized dynamically, as the analytic equality of the AdS and light-front expressions for electromagnetic and gravitational form factors of the composite states~\cite{Brodsky:2006uqa, Brodsky:2008pf}.

Therefore  in~\cite{deTeramond:2008ht}  Light-Front Holography was proposed as a first semi-classical approximation to QCD. The equations of motion, derived from the 5-dimensional AdS modified action,  are mapped to the wave equations of a system of two massless color-confined constituents in QCD quantized on the light front; the 5th coordinate in AdS$_5$, the holographic variable  $z$, is identified with the boost-invariant transverse  LF separation $\zeta$ (See the Appendix \ref{app}). In this approach,  the ``dictionary'' between the bound-state wave functions of the  classical five-dimensional theory and the four-dimensional theory quantized on the light front is fixed: for example, the bound state  consisting of a confined color-singlet quark and an antiquark with orbital and total angular momentum zero represents a pseudoscalar meson.

The implementation of the superconformal algebra in LFHQCD uniquely determines the form of the interaction. It also explains quantitatively the striking similarities between baryon and meson spectra~\cite{Dosch:2015nwa,Brodsky:2016yod,Dosch:2015bca,Dosch:2016zdv}  and fixes the modification of the AdS action. One predicts universal Regge slopes in  the radial and orbital quantum numbers $n$ and $L$ for both mesons and baryons. Only one mass parameter appears. A striking success of LFHQCD  is the prediction  of a massless pion~\cite{deTeramond:2008ht, Brodsky:2014yha} in the chiral limit. This is  a consequence of the implementation of the superconformal algebra, in contrast to the Goldstone mechanism in a theory with  a degenerate vacuum.

Isospin is not  introduced explicitly  in LFHQCD;  therefore massless particles do not only occur in the isovector, but also in the isoscalar  sector --a sector which has proven to be particularly challenging to  explain since the explicit breaking of the chiral $U(1)$ symmetry  of standard QCD is required.   We will show in this article that the spectroscopy of $\eta$- and $h$-meson states can  be quantitatively described by LFHQCD with the introduction of a single-parameter modification of the LF Hamiltonian.  A remarkable feature of the present approach is that  the numerical value of  the additional parameter coincides numerically with the LFHQCD confinement scale $\lambda$; this could point out to a deeper connection since in the chiral limit there is only one available scale in LFHQCD, the hadronic mass scale $\lambda$.

A consequence of the introduction of supersymmetry is the occurrence of tetraquarks as the second partners of the  baryons~\cite{Brodsky:2016yod,Nielsen:2018uyn,Zou:2018eam}. Since extra hadronic states appear to be particularly abundant in the isoscalar channels, it is promising to look specifically for possible tetraquarks in those channels. There is a tremendous literature on tetraquarks, and one can easily be lost in the possibilities. Supersymmetric LFHQCD, however, has the advantage that it makes quantitative and well defined predictions of the masses of these states. We will therefore perform a quantitative analysis of all of the tetraquark candidates among the isoscalar bosons. In this respect the present analysis is complementary to a  previous analysis~\cite{Nielsen:2018uyn} which gave a general overview over all possible tetraquark states predicted by supersymmetric LFHQCD. In this paper we give predictions for the masses of bosonic hadrons which can be specifically identified as tetraquarks within the extended LFHQCD scheme described here.

The paper is organized as follows: In Sec.~\ref{theo} we briefly review the main theoretical ingredients of LFHQCD.  We then
extend the  approach to  $\eta$- and $h$-mesons and discuss the implications for the Pauli principle on the quantum numbers of tetraquarks. In  Sec.~\ref{exp} we compare theory with experiment in all bosonic channels and predict the candidates which are most likely isoscalar tetraquarks. The possible relation of this modification of the LF Hamiltonian to chiral $U(1)$ breaking in standard QCD is briefly discussed in the last section, Sec.~\ref{sum}.

\section{Principal results of LFHQCD and supersymmetric LFHQCD \lb{theo}}

In this section we will review the main theoretical results for the hadron spectroscopy predicted by supersymmetric LFHQCD; a  more detailed treatment is given  in  Refs.~\cite{Dosch:2015nwa,Brodsky:2016yod,Brodsky:2014yha,Zou:2018eam}.  Some intermediate steps, which are most important for this paper, are given in Appendix \ref{app}.

As other holographic ``bottom-up" models, LFHQCD starts from an invariant action in a five-dimensional space with  the  metric of AdS$_5$.  Due to the maximal symmetry of the AdS$_5$ action, the corresponding 4-dimensional theory is invariant under the conformal group. This symmetry has to be broken by introducing a mass scale.  After such a modification, the resulting classical equations of motion of the 5-dimensional theory have the form of Hamiltonian equations for bound states of two massless quarks, where the $q \bar q $ interaction  is determined by the assumed modification of the  invariant AdS$_5$ action.   The unique form of the modified  AdS$_5$ action can in fact be completely determined by a symmetry principle: the resulting Hamiltonian of our semiclassical theory must  be contained within the superconformal algebra~\footnote{The  quantum field theory underlying the Maldacena conjecture is a superconformal theory.}.  This requirement  completely determines  the form of the color-confining $q \bar q$ interaction and consequently the modification of the AdS$_5$ action, both for mesons and baryons~\cite{deTeramond:2014asa,Dosch:2015nwa}. It also explains the observed approximate degeneracy between baryon and meson spectra and predicts the masses of the tetraquark states~\cite{Brodsky:2016yod}.

The resulting hadronic spectrum has the form of a supersymmetric 4-plet of $q \bar q$ mesons (M), quark+diquark baryons (B) and diquark+antidiquark tetraquarks (T), where in $SU(3)_C$ the diquark cluster has color $\bar 3_C$. The predicted hadron masses can, in the limit of massless quarks, be summarized in the following formul\ae~\cite{Brodsky:2016yod}:
\bea
 M_M^2 &=& 4 \lambda (n+L_M)+ 2 \lambda \,  \mathcal{S} \lb{spmes}, \\
 M_B^2 &=&4 \lambda(n+L_B+1) + 2 \lambda \,  \mathcal{S}  \lb{spbar}, \\
 M_T^2 &=&4 \lambda(n+L_T+1) + 2 \lambda \,  \mathcal{S}  \lb{sptet}.
\eea
Here $n$ denotes the radial excitation quantum number, $L_M$ denotes the LF orbital angular momentum between the quark and antiquark in the meson, $L_B$ that between the diquark cluster and the quark in the baryon, and $L_T$ that  between the two diquark clusters in the tetraquark.  $\mathcal{S}$ is for mesons the total quark spin, for baryons and tetraquarks the minimal possible quark spin of a diquark cluster inside the hadron. In supersymmetric LFHQCD only mesons with $J=L+\mathcal{S}$, $\mathcal{S}=0$ or $1$,  can be considered. The scale $\lambda$ is the only free constant of supersymmetric LFHQCD in the limit of massless quarks.

As mentioned above, the multiplets of supersymmetric LFHQCD contain only mesons, baryons and tetraquarks with total spin of the hadron  $J=L+\mathcal{S} $, $\mathcal{S}=0$ or 1.  But once the modification of the AdS$_5$ action is fixed by the superconformal constraints, one can apply  this modification also in normal  LFHQCD and derive the Hamiltonian for mesons with quark spin $\mathcal{S}=1$ and $J=L$ or $J=L-1$.   In this way we can also compare theory with the the observed mesons with $J^{PC}= 0^{++}$ and $1^{++}$. In this case the quantity $\mathcal{S}$ in \req{spmes} is replaced by $(J-L)$, see~\cite{deTeramond:2013it}.

In order to incorporate the effects of quark masses in LF theory, at least at lowest order, one can include the invariant mass term $\sum_i \frac{m_i^2}{x_i}$ to the LF Hamiltonian --the contribution of quark masses to the LF kinetic energy.   To first approximation, this leaves the confining LF potential unchanged. For a state containing $N$ quarks with masses $m_1, \dots, m_ N$ one then obtains the quadratic mass shift~\cite{Brodsky:2016yod}:
\be \label{DM2m}
\Delta M^2[m_1, \cdots, m_N] = \lambda^2\, \frac{\partial}{\partial \lambda} \log F,
\ee
with
\be
F[\lambda] = \int_0^1 dx_1 \cdots \int _0^1 dx_ N \, e^{- \frac{1}{\lambda}\left(\sum_{i=1}^N \frac{m_i^2}{x_i}\right)}
\,\delta\left(\sum_{i=1}^ N x_i -1\right) .
\ee

The quark mass corrections lead to a modified mass spectroscopy for the bosons:
\bea
M_M^2 &=& 4 \lambda (n+L_M)+ 2 \lambda \, (J-L) +\Delta M^2[m_1,m_2] \lb{spmes2},\\
M_T^2&=&4 \lambda(n+L_T+1) + 2 \lambda \, \mathcal{S}   +\Delta M^2[m_1,m_2,m_3,m_4]\lb{sptet2}.
\eea

The value of $\lambda$ has been fitted previously~\cite{Brodsky:2016yod} to the full hadron  spectrum with the result: $\sqrt{\lambda} = \kappa =0.523 \pm 0.025$ GeV. The effective masses of the light and strange quarks were determined in~\cite{Brodsky:2014yha} from
$m_\pi^2= \Delta M^2[m_q,m_q]$ and $m_K^2=\Delta M^2[m_s,m_q]$, which yields the values  $m_q=0.046$ and $m_s=0.357$ GeV. In Table~\ref{masscor} the numerical values of the mass corrections for non-strange and strange quark masses, according to \req{DM2m}, are collected. The numerical results for the boson masses $M_M$ and $M_T$ according to \req{spmes2} and \req{sptet2} are given in Table~\ref{tabtest}.

\begin{table}
\caption{\lb{masscor} Mass corrections according to \req{DM2m} in GeV$^2$.}
\begin{tabular}{ll} \hline \hline
$\Delta M^2[m_q,m_q]= (0.14)^2$&\\$
\Delta M^2[m_s,m_s]= (0.773)^2$ \\
$ \Delta M^2[m_q,m_q,m_q,m_q]=(0.344)^2$ \\
$\Delta  M^2[m_q,m_s,m_q,m_s]=(0.959)^2$\\
 $\Delta M^2[m_s,m_s,m_s,m_s]=(1.53)^2$&\\
 \hline \hline
\end{tabular}
\end{table}

\begin{table}
\caption{\lb{tabtest}  Boson masses according to \req{spmes2} and \req{sptet2} in GeV.}
\begin{tabular}{c|ccccc} \hline \hline
&\multicolumn{5}{c}{$M$ [GeV]}\\
Quant. num. &\multicolumn{5}{c}{Quark content}\\
$n+L+\frac{J-L}{2}$&$  ~ \bar q q_0$&$ ~ \bar s s_0$&$ ~ \overline{qq}\,qq$&$ ~ \overline{qs}\,qs$&$ ~ \overline{ss}\,ss$\\  \hline
    0 & 0.14 & 0.77 & 1.1 & 1.42 & 1.85 \cr 1/2 & 0.753 & 1.07 & 1.33 & 1.6 & 2. \cr 1 & 1.06 & 1.3 & 1.52 &
   1.76 & 2.13 \cr  3/2 & 1.29 & 1.49 & 1.69 & 1.91 & 2.25 \cr 2 & 1.49 & 1.67 & 1.84 & 2.05 & 2.37 \cr  5/2 &
   1.66 & 1.82 & 1.99 & 2.18 & 2.48 \cr 3 & 1.82 & 1.97 & 2.12 & 2.3 & 2.59 \cr  7/2 & 1.96 & 2.1 & 2.24 &
   2.42 & 2.7 \cr 4 & 2.1 & 2.23 & 2.36 & 2.53 & 2.79 \cr  9/2 & 2.22 & 2.35 & 2.48 & 2.63 & 2.89 \cr 5 &
   2.34 & 2.46 & 2.58 & 2.74 & 2.98\\
   \hline\hline
\end{tabular}
\end{table}

\subsection{$\eta$-and $h$ mesons \lb{iso}}

 As can be seen from \req{spmes} the ground states ($n=0$) of  pseudoscalar mesons with angular momentum $J=L=0$ in LFHQCD  have zero mass in the limit of massless quarks.  This prediction of a massless pseudoscalar meson $q \bar q$ bound state is a remarkable  success of LFHQCD for the isovector channel.  But since LFHQCD does not treat flavor explicitly,  this result also applies to the isoscalar channel.  In this case, however, the least massive observed hadron is the $\eta$ meson, which has a mass of 0.548 GeV, much heavier than the pion.

In standard QCD the difference between the isovector and isoscalar sector is generally attributed to a hard breaking of the chiral $U(1)$ symmetry of the classical QCD Lagrangian by nonperturbative effects~\cite{Kogut:1973ab,tHooft:1976rip,Witten:1979vv,Veneziano:1979ec,Veneziano:1980xs,DiVecchia:1980yfw,Engelhardt:2000wc,Alkofer:2008et,Kataev:1981aw}.
Since flavor is not treated explicitly  in LFHQCD we will treat that breaking  phenomenologically by ensuring that the lowest $I = 0, J = \mathcal{S} = 0$ meson has the correct mass $m_\eta = 0.548$ GeV~\cite{Tanabashi:2018oca}. This can be achieved by adding to the LF Hamiltonian the  constant term $\Delta_\eta^2 \,\lambda\, \delta_{\mathcal{S} 0}\delta_{I 0}$, with
\be \lb{Del2}
 \Delta_\eta^2 = \frac{1}{\lambda} \left(M_\eta^2-\Delta M^2[m_q,m_q]\right)
 = \frac{1}{\lambda} \left(M_\eta^2 - M_\pi^2 \right),
\ee
a dimensionless quantity with the value $ \Delta_\eta^2 = 1.03\pm0.08$. This additional term encodes a hard chiral $U(1)$ breaking in standard QCD. In the isoscalar vector channel current conservation in QCD forbids an anomaly; accordingly,  the effective light front Hamiltonian should not be modified in this sector. The scale associated with the $\eta$ mass, namely $\lambda \Delta_\eta^2$, numerically nearly coincides with  the confinement scale $ \lambda = (0.523~\pm~0.025 \mbox{ GeV})^2$~\cite{Brodsky:2016yod}.  It is therefore tempting to speculate there is a deeper connection behind this numerical equality and that the two scales are indeed the same since there is a unique mass scale in LFHQCD. We will, however, not discuss this issue further in this article.

The resulting mass formul\ae  \, for mesons and tetraquarks  in the isoscalar sector are:
\bea
M_M^2  &=& 4 \lambda (n + L_M) + \Delta M^2[m_1,m_2]+
2  \lambda \mathcal{S}+ \lambda \Delta^2_\eta ,
\lb{M}\\
M_T^2 &=&  4 \lambda (n +  L_T + 1)  + \Delta M^2[m_1,m_2,m_3,m_4]  +
2 \lambda \mathcal{S}+ \lambda \Delta^2_\eta , \lb {T}
\eea
where $ \Delta^2_\eta$ is given by  \req{Del2}. The satisfactory comparison with experiment will be given in Sect. \ref{exp}.

\subsection{Isospin and spin of tetraquarks \lb{tetra}}

The two quarks of a diquark cluster in a tetraquark are antisymmetric in color;  the spin-statistics theorem therefore demands that they are symmetric in the remaining quantum numbers.  Therefore a diquark cluster with specific isospin and relative orbital angular momentum zero must either have isospin and spin both equal to zero, or both equal to 1. This implies that the lowest lying tetraquark with isospin 0 must have total angular momentum $J=0$ and the one with isospin 1 must have $J=1$;  as a result, the squared masses of the two states differ by $2 \lambda$.

These arguments do not apply to tetraquarks containing constituents which  are not related by isospin symmetry.  The lowest state consisting of the type $(\overline{qs} qs)$ is isospin degenerate and has in supersymmetric LFHQCD the mass 1.42 GeV. Since diquark clusters are bosons, the parity and C-parity of a tetraquark with $\mathcal{S}=0$ is $(-1)^L$.  For a tetraquark with $\mathcal{S}=1$ both C-parities are possible.

For a tetraquark consisting only of strange and antistrange quarks,  the spin-statistics theorem requires that the total spin of each diquark cluster  is 1. Therefore such tetraquarks are not predicted in the scheme of supersymmetric LFHQCD. We  will therefore restrict our general discussion to tetraquarks, where only one diquark cluster has spin 1.  But at  least tentatively, we suggest that the higher lying states of $f_2$ mesons might contain admixtures of tetraquarks $ (\overline{ss}ss)$ and extend  \req{sptet2} to  $\mathcal{S}=2$. This is motivated by the appearance of two $\phi$ mesons in the decay channels of these mesons.

\section{Comparison with experiment\lb{exp}}

The augmented LFHQCD theory presented here contains four parameters.  Three of them, the scale $\sqrt{\lambda}=\kappa=0.523$ GeV and the quark masses $m_q=0.045, \, m_s=0.357$ GeV, are taken from previous analyses of the hadron spectrum \cite{Brodsky:2016yod}.
 For this analysis  the shift term $\Delta^2_\eta = \lambda_\eta \,\delta_{\mathcal{S}0}\,\delta_{I0}$ has been introduced in the LF Hamiltonian for mesons \req{M} and tetraquarks  \req{T} in the isoscalar sector. As discussed above,  the scale $\Delta_\eta^2\,\lambda$ is fixed by the $\eta$ mass, $\Delta_\eta^2 \,\lambda\equiv M_\eta^2 -  M_\pi^2  \simeq  \lambda$.

\subsection{$\eta$ and $h$ mesons \lb{isoex}}

As can be seen from Fig.~\req{md},  the mass difference $\sqrt{M_{\eta'}^2-M_\eta^2}$  agrees reasonably well with the corresponding mass differences between other hadrons  $x$ and $y$ with the same external quantum numbers. This quantity  is determined by the mass difference between the strange and the  light quarks:  two light quarks are replaced by two strange quarks in baryons, and for mesons the comparison is made for isoscalar mesons with the same quantum numbers. If the hadrons $x$ and $y$ are mesons one can conclude from their decays that  the heavier meson $x$  is predominantly an $\bar s-s$ pair, whereas the lighter hadron $y$ is a $\bar q-q$ pair, both with the same quark spin $\mathcal{S}$. If the hadrons $x$ and $y$ are baryons, then they both have the same quantum numbers $J^P$, but the  strangeness of the heavier one is larger by two units; {\it i.e.}, the $q-q$ diquark cluster of  the lighter baryon $y$  is replaced by an $s-s$ cluster in the heavier baryon $x$.  The mean value of the mass difference $M_d=\sqrt{M_x^2-M_y^2}$ is  0.84 GeV with a standard deviation of 0.09 GeV;  the theoretically predicted value is
$\sqrt{\Delta^2[m_s,m_s]-\Delta^2[m_q,m_q]}=  0.76$ GeV  (see Table \req{masscor}).

Therefore in the present   analysis the  $\eta$-$\eta'$ mass difference, sometimes referred to as the $\eta - \eta'$ puzzle, is determined by the strange quark mass contribution
\be \lb{etap}
M_{\eta'}^2 =\Delta_{\eta}^2 \lambda + \Delta M^2[m_s, m_s] =M_\eta^2-\Delta^2[m_q,m_q] + \Delta^2[m_s,m_s].
\ee
Using the results of Table \req{masscor}, Eq.~\req{etap} leads to $M_{\eta'} = 0.937$ GeV,  in good agreement with the experimental value $M_{\eta'} = 0.958 \pm 0.06$ GeV~\cite{Tanabashi:2018oca}.
\begin{figure}
\bec
\includegraphics[width=10cm]{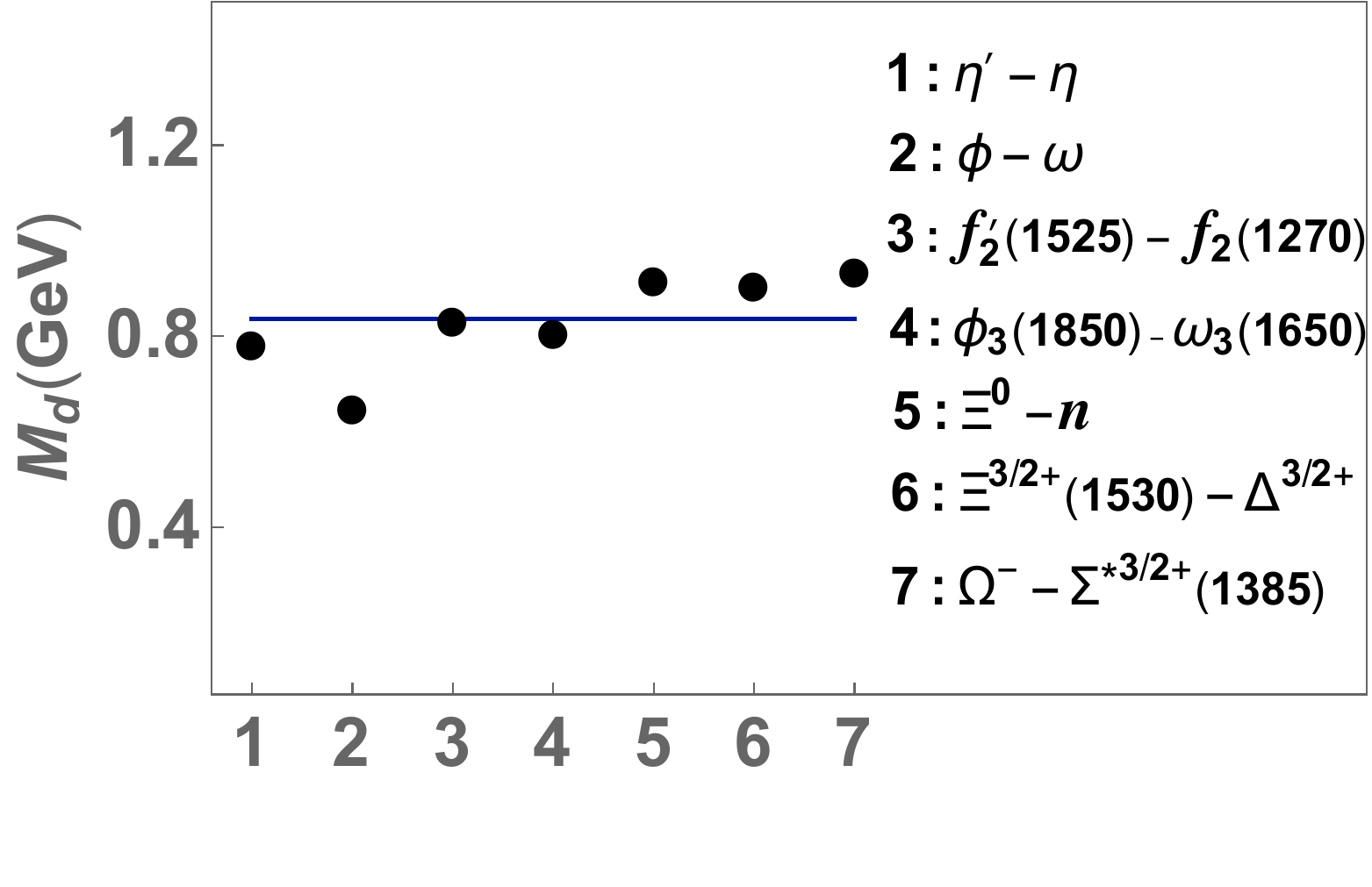}
\enc
\caption{ The mass differences  $M_d=\sqrt{M_x^2-M_y^2}$ between states with the same $J^{PC}$  and $J^P$ respectively.}
\lb{md}
\end{figure}

In Fig.~\ref{etafig} the theoretical trajectories for the $\eta$, $\eta'$ and $h$ mesons and their radial and orbital excitations are shown. Since not very many $\eta$ mesons, and even fewer $h$ mesons, have been confirmed experimentally, we have also considered unconfirmed states  listed in~\cite{Tanabashi:2018oca}.    As can be seen from the figure, the agreement between theory and experiment is for the ground states indeed satisfactory, particularly, the $\eta-\eta'$ mass difference is explained. As in other quark-spin zero cases the agreement between the simple theory and experiment is not so good for radial excitations. Presumably mixing plays some role there.  For a more detailed discussion, see the next subsection.

\begin{figure}
\begin{center}
\includegraphics[width=10cm]{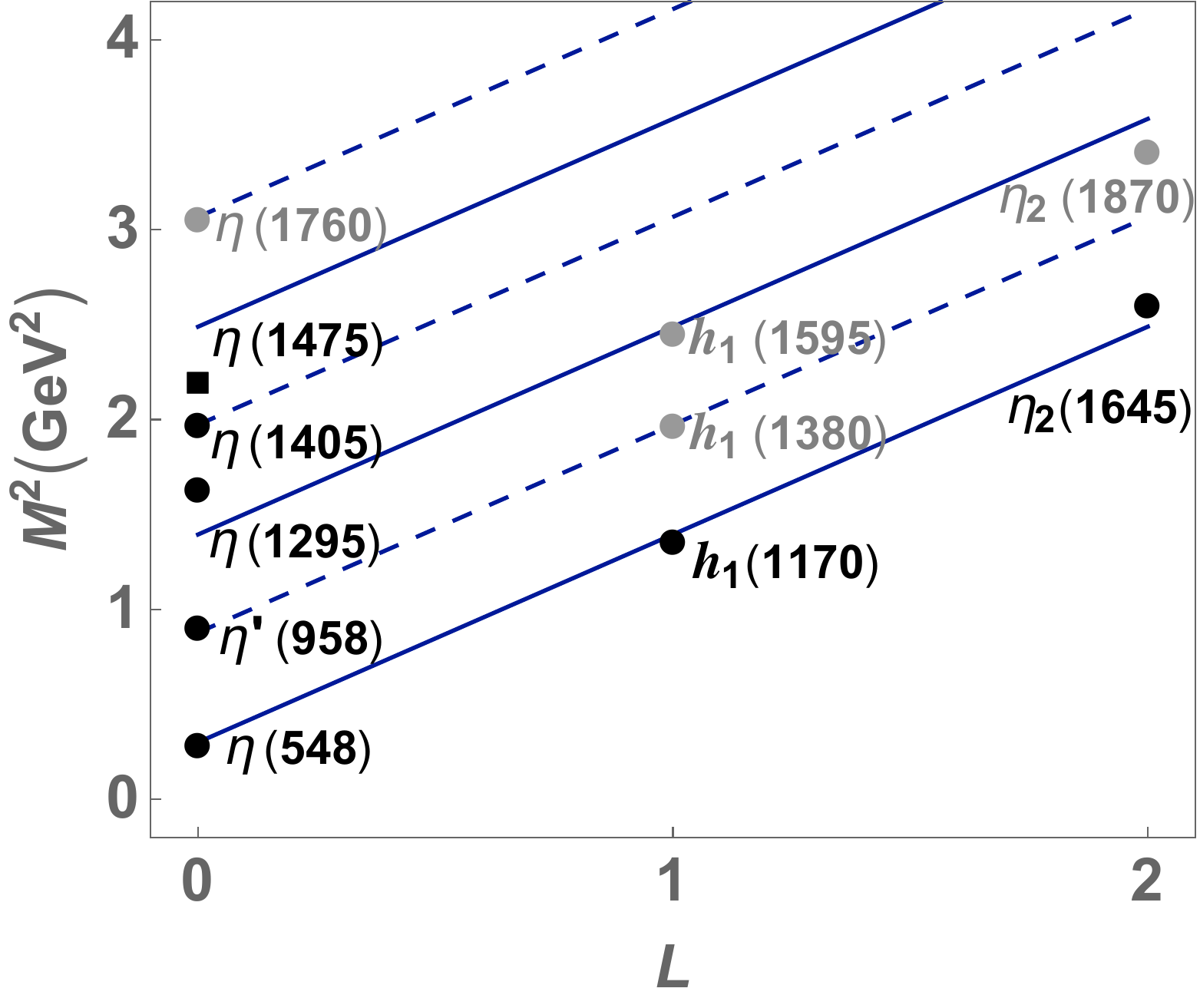}
\caption{\lb{etafig}The $I=0,\, J=L $ and $C=(-1)^L$ states. The theoretical  predictions  for the meson trajectories are from  \req{M}: Continuous lines $\eta$ families, dashed lines $\eta'$ families. Data  (filled circles and square) are from PDG~\cite{Tanabashi:2018oca}, unconfirmed states are in gray.}
\end{center}
\end{figure}

\subsection{General comparison of isoscalar bosons}

In Table~\ref{tabthex} we show all of the confirmed isoscalar bosons. In the first 4 columns we show the experimental results;  the letter  $d$ in the column ``Decay" indicates that  decay channels with open or hidden strangeness are dominant,  and $ss$ indicates that there are decay channels with four-fold hidden strangeness, such as the $\phi\,\phi$, for the $\eta'$ decay (see the discussion below).  In the four columns listed under  ``Theory" we show  below $M_{nL,J-L}$ and $T_{nL,J-L}$ the quark content, the radial excitation number $n$, the light front angular momentum $ L$, and the difference $J-L$, as indices.  For the $\eta$ like mesons with $J^{PC}=0^{-+}$ or $1^{+-}$ we denote by an upper index $^\eta$ the fact that the shift given by $\Delta^2_\eta \lambda$ , see {Sect.}~\ref{theo}, has to be performed;  the theoretical masses are calculated according to (\ref{spmes2}, \ref{sptet2}) and (\ref{M}, \ref{T}), respectively.

\begin{table}
\caption{\lb{tabthex} All confirmed light isoscalar bosonic states. The data are from the PDG~\cite{Tanabashi:2018oca}, and the theoretical results from  (\ref{spmes2}, \ref{sptet2})  and (\ref{M}, \ref{T}).  The label $d$ signifies that channels with open or hidden strangeness are dominant; for the $\eta'$ decay see the note in the text.  If a state with $I=1$, but equal residual quantum numbers occurs, it is indicated in the last column.
 For further explanation see the text.}
\begin{tabular}{cccc|cccc|c} \hline \hline
\multicolumn{4}{c|}{Experiment} & \multicolumn{4}{c|}{Theory}&\\
$J^{PC}$&Name&M ~[MeV]& Decay&\multicolumn{2}{c}{Meson}& \multicolumn{2}{c|}{Tetraquark}&$I=1$\ partner\\
&& & &$M_{nL,J-L}$&[GeV] &$T_{nL,\mathcal{S}}$&[GeV]& $$\\
\hline
$0^{-+}$&$\eta$&$548$&&$\bar q q^\eta_{000}$&$(0.548)$&&&\\
$0^{-+}$&$\eta'(958)$&$958$&  &$\bar ss^\eta_{000}$&$0.94$&&\\
$0^{-+}$&$\eta(1295)$&$ 1294 \pm 4$&&$\bar qq^\eta_{100}$&$1.18$&&\\
$0^{-+}$&$\eta(1405)$&$1409  \pm 2$&&$\bar ss^\eta_{100}$&$1.41$&&&\\
$0^{-+}$&$\eta(1475)$&$1476  \pm 4$&d&$\bar ss^\eta_{?00}$&&&\\
\hline
$0^{++}$&$f_0(500)$&$475\pm75$&&&&&&\\
$0^{++}$&$f_0(980)$&$990\pm20$&&$\bar q q_{01-1}$&$0.75$&
$(\overline{qq} qq)_{000}$&1.10& $a_0(980)$ \\
$0^{++}$&$f_0(1370)$&$1350\pm150$&&$\bar q q_{11-1}$&$1.29$&&$$&$$\\
$0^{++}$&$f_0(1500)$&$1504\pm6$&&&&$(\overline{qq} qq)_{100}$&1.52&\\
$0^{++}$&$f_0(1710)$&$1723\pm6$&d&$\bar s s_{21-1}$&$1.82$&&\\
\hline
$1^{--}$&$\omega$&$783$&&$\bar qq_{001}$&$0.75$&&&$\rho$\\
$1^{--}$&$\omega(1420)$&$1425\pm 25$&&$\bar q q_{101}$&$1.29$&$(\overline{qq} qq)_{010}$&$1.52$&$\rho(1450)$\\
$1^{--}$&$\omega(1650)$&$1670\pm  30$&&$\bar qq_{201}$&$1.66$&&&\\
$1^{--}$&$\phi$&$1019$&d&$\bar s s_{001}$&$1.07$&&\\
$1^{--}$&$\phi(1680)$&$1680\pm 20$&d&$\bar ss_{101}$&$1.49$&$(\overline{sq} sq)_{010}$&$1.76$&\\
$1^{--}$&$\phi(2170)$&$2188\pm 10$&  &$\bar ss_{301}$&$2.10$&$$&$$&\\
\hline
{$1^{+-}$}&$h_1(1170)$&{$1170\pm 20 $}&&$\bar qq^\eta_{010}$&1.18&&\\
\hline
$1^{++}$&$f_1(1285)$&$1282$&&$\bar q q_{110}$&$1.49$&&&$a_1(1260)$\\
$1^{++}$&$f_1(1420)$&$1426$&d&$\bar s s_{010}$&$1.30$&$(\overline{sq} sq)_{001}$&{1.60}\\
\hline
$2^{-+}$&$\eta_2(1645)$&$1617\pm 5$&&$\bar q q^\eta_{020}$&$1.58$&&&\\
\hline
$2^{++}$&$f_2(1270)$&$1276$&&$\bar q q_{011}$&$1.29$&&&$a_2(1320)$\\
$2^{++}$&$f_2'(1525)$&$1525 \pm 5 $&d&$\bar ss_{011}$&$1.49$&&&\\
$2^{++}$&$f_2(1950)$&$1944 \pm 12$&&$\bar s s_{111}$&$1.82$&&&\\
$2^{++}$&$f_2(2010)$&$2011 \pm 70$&ss&$\bar s s_{211}$&$2.10$&$(\overline{ss} ss)_{002}$&$2.13$&\\
$2^{++}$&$f_2(2300)$&$2297 \pm 28$&ss&$\bar s s_{311}$&$2.35$&$$&$$&\\
$2^{++}$&$f_2(2340)$&$2345 \pm 50$&ss&$$&$$&$(\overline{ss} ss)_{102}$&$2.37$\\
\hline
$3^{--}$&$\omega_3({ 1670})$&$1667\pm 4$&&$\bar qq_{021}$&$1.66$&&&$\rho_3(1690)$\\
$3^{--}$&$\phi_3(1850)$&$1854\pm 7$&&$\bar ss_{021}$&$1.82$&&&\\
\hline
$4^{++}$&$f_4(2050)$&$2018\pm 11$&&$\bar qq_{031}$&$1.96$&&&$a_4(2040)$\\
\hline\hline
\end{tabular}
\end{table}

We will leave out from this comparison the extremely broad  $f_0(500)$ which will be discussed later.  The overall fit from theory to experiment is satisfactory -- the standard deviation (SD) between  theory and experiment is SD =  93 MeV, well inside the model uncertainty of $\approx 100$ MeV,  as  expected from the $N_C \rightarrow\infty$ expansion~\cite{tHooft:1974}. The discrepancy between theory and experiment is no more than 3 standard deviations for any of the considered 27 states. Therefore we accept only as probable a tetraquark assignment for states where the difference between the experimental masses and  LF holographic predictions is less than 3 SD $\approx 280$ MeV.

We now start a detailed discussion of the states:

\paragraph{\bf $\eta$ and $h$ mesons, $J^{PC}=0^{-+}, \,1^{+-},\,2^{-+}$}
 In our approach the $\eta$ is predominantly $\bar q q$ and the $\eta'$ is predominantly $\bar s s$. Since the $\eta$ is slightly below the 4$\pi$ threshold and the $\eta'$ below the $ K \overline K$ threshold we cannot test this assignment by the decays, but the width of the $\eta'(958)$ meson is very small (196 keV). The dominant decay $\eta \pi \pi$ is  not forbidden by any selection rule or spin and statistics. This is an additional  argument that $\eta$ and $\eta'$ have a different quark content, with the $\eta'$ having an important hidden strangeness. As mentioned in sect. \ref{iso}, the agreement between theory and experiment is satisfactory. Presumably, the $\eta(1295),\,\eta(1405)$ and the $\eta(1475)$ are mixed states of the $\eta(2S)$, $\eta'(2S)$ and $\eta(3S)$.

\paragraph{\bf $f_0$ states} The assignment of  $f_0(500)$ and the $f_0(980)$ as members of a nonet of tetraquarks, with and without hidden strangeness, appears very plausible~\cite{Nielsen:2018uyn,Hooft:2008we,Black:1999yz,Jaffe:2003sg,Jaffe:1976ig,Maiani:2004uc}; however, quantitative predictions from LFHQCD do not add support to this assignment: The lightest tetraquark consisting of non-strange quarks, $(\overline{qq}qq)_{000}$ has a mass of 1.10 GeV, compatible with the $f_0(980)$, but the lightest tetraquark with hidden strangeness, $(\overline{sq}sq)_{000}$ has a mass of 1.42 GeV.  The conventional meson in a $q \bar q\left(^3P_0\right)$ state, $\bar q q_{01-1}$, has in LFHQCD the mass 0.75 GeV which is also in this mass range. Unfortunately, the quantitative predictions from LFHQCD do not contribute to the solution of this interesting situation.

The other $f_0$ states fit very well into the LFHQCD theoretical scheme. The $f_0(1500)$ is a very good candidate for a radially excited tetraquark; its mass fits nicely, and its baryonic  partner is the Roper resonance $N(1440)$, see Table \ref{tabtetra}.

\paragraph{\bf $f_1$ states}  The situation is similar as for the $f_0(980)$.  The $f_1(1282)$ is heavier than the $^1P_1$ ground state (1.06 GeV) and lighter than its first
radial excitation (1.49 GeV), there is also no tetraquark with fitting mass and the occurrence of the near degenerate
isospin partner $a_1(1260)$ make an interpretation as a meson state plausible anyhow.  An interpretation of the $f_1(128
2)$
and $a_1(1260)$ as tetraquarks with $\mathcal{S}=1$ and hidden strange $ (\overline{sq}sq)_{001}$~\cite{Nielsen:2018uyn} is
not supported  by the quantitative LFHQCD analysis, the discrepancy between the theoretical mass value for this assignment and experiment is 315 MeV $\sim$ 3.6 SD.

\paragraph{\color{black}\bf $f_2$ states} Our quark assignment in Table~\ref{tabthex} for the $f_2(1270)$ and the $f'_2(1525)$ is compatible with a recent lattice computation above threshold, where the lighter state couples predominantly to $\pi \pi$ and the heavier to $K \overline K$~\cite{Briceno:2017qmb}. On the other hand, the profusion of $f_2$ states makes the occurrence of exotic states very plausible.  Furthermore the occurrence of two $\phi$-mesons in the decay channels of the heavier $f_2$ mesons strongly suggest the admixture of a tetraquark consisting of  two strange and two anti-strange quarks.  As mentioned in Sec.~\ref{iso}, a diquark cluster with strangeness $\pm2$ must have spin 1, so it is plausible that these states contribute  to mesons with $J=2$.  Extending \req{sptet2} also to diquarks with spin 2, we obtain for the lightest tetraquarks with quark content  $(\overline{ss}ss)_{002}$ the mass $M_T=2.13$ GeV, and for the $(\overline{ss}ss)_{102}$  configuration $M_T=2.37$ GeV; they are just in the mass range of those states which decay into two $\phi$ mesons. We therefore propose that the $f_2(2010), f_2(2300)$ and $f_2(2340)$ are mixtures of normal meson states and tetraquark states, as indicated in Table~\ref{tabthex}. These tetraquarks with aligned spin have positive parity and charge parity $ P = C = 1$.

\begin{table}
\caption{\lb{tabtetra} A compilation of the tetraquark states and their partners. States omitted from the summary table of PDG are marked by a question mark $^?$. The indices of the quark content denote $n,L,\mathcal{S}$, the radial excitation, LF momentum and the quark or diquark spin. The differences of the theoretical masses, M$_{theo}$ shows the amount of SUSY breaking due to the additional mass terms. Assignments with remark $C$ do not explain the occurrence of isovector states with the same residual quantum numbers and similar masses.}
\begin{tabular}{ccc|ccc|ccc|c}
\hline\hline
\multicolumn{3}{c|}{Tetraquark}&\multicolumn{3}{c|}{Baryon}&\multicolumn{3}{c|}{Meson}&Remark\\
{\footnotesize Name}& {\footnotesize quark content}&{\footnotesize M$_{theo}$}&{\footnotesize Name}& {\footnotesize quark content}&{\footnotesize M$_{theo}$}&{\footnotesize Name}& {\footnotesize quark content}&{\footnotesize M$_{theo}$}&\\
\hline
$f_0(980)$&$(\overline{qq}qq)_{000}$& 1.10&$N(940)$&$(qqq)_{000}$&1.07&$h_1(1170)$&$\bar q q^\eta_{010}$& 1.18&C\\
$\omega(1420)$&$(\overline{qq}qq)_{010}$& 1.52&$N^{\frac{1}{2}-}(1535)$&$(qqq)_{010}$&1.50&$\eta_2(1645)$&$\bar q q^\eta_{020}$& 1.58&C\\
$f_0(1500)$&$(\overline{qq}qq)_{100}$& 1.52&$N^{\frac{1}{2}+}(1440)$&$(qqq)_{100}$&1.50&$h_1(1595)^?$&$\bar q q^\eta_{110}$& 1.58&\\
$f_1(1420)$&$(\overline{sq}sq)_{001}$& 1.6&$\Xi^{\frac{3}{2}+}(1530)$&$(ssq)_{001}$&1.50&$f_2'(1525)$&$\bar s s_{011}$& 1.49&\\
$\phi(1680)$&$(\overline{sq}sq)_{010}$& 1.76&$\Xi^*(1690)^?$&$(ssq)_{010}$&1.67&$\eta_2(1870)^?$&$\bar s s^\eta_{020}$& 1.75&\\
\hline\hline
\end{tabular}
\end{table}

In Table~\ref{tabtetra} we have  listed those  tetraquark states which fit quantitatively into the scheme of supersymmetric LFHQCD. The masses of the first two,  the $f_0(980)$ and the $\omega(1420)$, fit reasonably well into the LFHQCD scheme;  although the isospin degeneracy with the $a_0(980)$ and the $\rho(1440)$ cannot be explained by this assignment. The remaining four states are, however, very probably tetraquark states. Note that the additional mass correction term leads to a breaking of the supersymmetry, as can be seen from the different theoretical masses inside one supermultiplet.

\section{Summary and discussion \lb{sum}}

As has been shown in Sec.~\ref{iso}, the addition of a constant term to  the $I=0,\, \mathcal{S}=0$ bosonic sector of the supersymmetric
light front holographic Hamiltonian provides an explanation of the entire $\eta-h$ spectrum. This additional term, which
breaks the supersymmetry of the spectra, plays the role of a direct chiral $U(1)$ breaking term in
effective chiral representations of standard QCD. In fact, at the classical level, the QCD Lagrangian with massless {\it up} and  {\it down} quarks is  invariant under $U_L(2) \otimes U_R(2)$  transformations. In conventional effective chiral theory, $SU_L \otimes SU_R(2)$ is broken spontaneously and leads to an isovector of Goldstone particles which are identified with the $\pi$ mesons. Since the isoscalar pseudoscalar meson, the $\eta$-meson, is considerably heavier  than $\sqrt{3}\, M_\pi \approx 237$ MeV~\cite{Weinberg:1975ui} there is apparently  no  Goldstone boson of the remaining chiral $U(1)$ symmetry and it  is most probably broken directly,  for instance by instanton solutions or other nonperturbative effects, which modify the effective QCD Hamiltonian~\cite{Kogut:1973ab,tHooft:1976rip,Witten:1979vv,Veneziano:1979ec,Veneziano:1980xs,DiVecchia:1980yfw,Engelhardt:2000wc,Alkofer:2008et,Kataev:1981aw}.

In supersymmetric LFHQCD,  the implementation of the superconformal algebra is the origin of the vanishing mass of mesons with $L=0, \, \mathcal{S}=0$, since it predicts a constant term $-2 \lambda$ in the LF potential which exactly cancels the LF  kinetic energy.  Since the lowest meson state has no supersymmetric partners~\cite{Dosch:2015nwa}, it plays the role of a zero-energy non-degenerate ground state.
This occurrence of massless mesons  is conventionally associated with the spontaneous breaking of the chiral $SU(2)$ symmetry of QCD based on effective meson fields.

 But if the chiral $U(1)$ symmetry is broken directly, also the supersymmetry of LFHQCD needs a hard breaking. Such a symmertry breaking term can be most easily incorporated into  light front holographic QCD by adding to the LF potential  $U_{AdS}(\zeta)$, see \req{eqads}, for the  bosonic channels  the constant term $\lambda \Delta_\eta^2\, \delta_{\mathcal{S} 0}\delta_{I 0} \approx \lambda \, \delta_{\mathcal{S} 0}\delta_{I 0}$~\req{Del2}. It is remarkable that this simple modification of the  potential, which is determined by the eta mass, see \req{Del2},  can explain the full $\eta$ and $h$ meson spectrum, including in particular the $\eta'- \eta $ mass difference, as shown in Fig.~\ref{etafig}. It should be noted that also in more standard analytical approaches to the $U(1)$ problem~\cite{Christos:1984tu} at least one additional parameter is necessary, be it an instanton cutoff~\cite{Belavin:1975fg,tHooft:1976rip} or  mixing angles~\cite{Escribano:2005qq}. In the channels with quark spin $\mathcal{S}=1$ no symmetry-breaking term is present and no modification of supersymmetric  LFHQCD is necessary.

We have also considered states with $J \neq L+\mathcal{S}$, which are not members of super-multiplets; their masses are predicted by LFHQCD,  given the  modification of the AdS action which  is determined by the implementation of the superconformal algebra.  The theoretical  mass predictions  are collected in  Fig. \ref{etafig} and Table~\ref{tabthex}.

The quantitative discussion of tetraquarks in this investigation is complementary to that of Ref.~\cite{Nielsen:2018uyn}, where a general qualitative overview of possible supermultiplets in all channels was given.  In this paper we have concentrated on the possible tetraquark states,  the mass of which agrees within the model accuracy of $\approx 100$ MeV  with the theoretical  predictions.  Candidates fulfilling these criteria are collected in Table~\ref{tabtetra}, together with their partners. The most convincing candidate is the $f_0(1500)$.  The $f_0(1500)$ has also been considered as a candidate for a glueball, see {\it e.g.},~\cite{Close:2000yk,Amsler:1995td};  it should be noted,  however,  that in
 LFHQCD there is no sign of valence gluons, and the identification of the $f_0(1500)$ as a tetraquark is quantitatively very convincing.

In this article we have concentrated on the new insights that  LFHQCD brings to the symmetry breaking mechanisms of standard QCD. We have also emphasized the remarkable numerical coincidence of the magnitude of the ``symmetry breaking" term $\lambda \Delta \eta^2$ with the value of the confinement scale  $\lambda$ of LFHQCD. In fact, the results described in this article provide a new correspondence between the chiral symmetries of standard QCD and LFHQCD: In  the conventional approach, the occurrence of a massless pseudoscalar particle in the massless quark limit is assumed to be a consequence of the Goldstone theorem,  whereas in supersymmetric LFHQCD, the massless of the composite  $q \bar q$ pion bound state  is an explicit feature of the lowest eigenvalue of the LF Hamiltonian. Superconformal LFHQCD thus leads to a novel alternative mechanism for the  origin of  massless composite pseudoscalar mesons.

The explicit breaking of the $U(1)$ classical symmetry, whatever its reason in standard QCD, can be incorporated into LFHQCD by adding a constant to the superconformal potential. The confining term $\lambda \zeta^2$, which is a consequence of the conformal algebra alone~\cite{Brodsky:2013ar}, is not affected. The remarkable fact that the symmetry breaking term $ \lambda \Delta_\eta^2$ is equal within errors to the only scale available in LFHQCD  in the chiral  limit, \req{Del2}, indicates a deep connection of universal mass scales in QCD.

\section{Acknowledgements}

One of us (HGD) wants to thank the Chinese Academy of  Sciences  and especially Prof.  Pengming  Zhang  for the warm hospitality extended to him at the Institute of Modern Physics, Lanzhou (Gansu). This work is supported in part by the Department of Energy, Contract DE--AC02--76SF00515 and by the National Natural Science Foundation of China (Grants No.~11575254 and 11805242).

\appendix

\section{Summary of LFHQCD results \lb{app}}

In this appendix we summarize the relevant equations from LFHQCD which underlie the theoretical foundations of this paper.
The equations of motion for a meson with arbitrary spin-$J$, represented by a fully symmetric tensor field of rank-$J$ in 5 dimensional AdS space, follows from an AdS$_5$ action  with the soft-wall dilaton term $e^{\varphi(z)}$~\cite{deTeramond:2013it}:
\be \lb{eqads}
\left( - \frac{d^2}{d z^2} + \frac{4 L^2_{AdS} -1}{4 z^2} +U_{AdS}(z)\right) \Phi_J(z,q)= q^2  \Phi_J(z,q).
\ee
Here $z$ is the fifth coordinate of AdS$_5$, the holographic variable in the 5-dimensional space, and $q^2$ the momentum in the 4-dimensional (physical) spacetime. The potential $U_{AdS}(z)$ depends on the dilaton profile  $\varphi(z)$:
\be \lb{uads}
U_{AdS}(z)= \half \varphi''(z) + \textstyle{\frac{1}{4}} \left( \varphi'(z)\right)^2 + \frac{2 J-3}{2\, \zeta}\, \varphi'(z).
\ee
The quantity $L^2_{AdS} $ is determined by $J$ and the dimensionless product of the AdS-mass $\mu$ with the the space curvature $R$: $L^2_{AdS}=(\mu R)^2+  (2 - J)^2$.

The form of \req{eqads} is that of a bound-state equation for a hadron consisting of two massless constituents in light-front quantization. The holographic variable $z$ is identified with the boost-invariant LF variable $\zeta = \sqrt{x(1-x)} \, b_\perp$, where $x$ is the longitudinal momentum fraction of one of the quark constituents and $b_\perp$ is the transverse separation of the constituents (quarks or quark clusters) in the transverse plane.  The LF angular momentum $L$ is identified with the quantity $L_{AdS}$  and the light-front potential $U(\zeta)$ is therefore determined by \req{uads}.

If one implements superconformal algebra~\cite{deTeramond:2014asa, Dosch:2015nwa} by requiring  the LF Hamiltonian to be a  a superposition of the generators of the superconformal algebra following~\cite{deAlfaro:1976vlx, Fubini:1984hf}, the form of the LF potential is completely fixed to
\be \lb{ulf}
U(\zeta)= \lambda^2 \,\zeta^2 + 2 \lambda (L-1) ,
\ee
which leads to eigenvalues
\be  \lb{M2la}
M^2 = \vert \lambda \vert ( 4n + 2 L + 2) + 2 \lambda (L - 1).
\ee

The potential derived from the implementation of the superconformal algebra is only compatible with the holographic approach if the  dilaton profile $\phi(z) = \lambda z^2$ and holds for  mesons with $J=L$.  It is remarkable that this choice of maximal symmetry breaking is the one which had been chosen before in the soft wall model~\cite{{Karch:2006pv}} for purely phenomenological reasons,  namely to generate linear Regge trajectories for mesons. A zero mass state only occurs if  the sign of $\lambda$ is  positive:
\be  \lb{M2lap}
M^2 = 4 \lambda ( n + L),
\ee
and  therefore the lowest meson state has no baryon partner~\cite{Dosch:2015nwa, Witten:1981nf}.
One sees immediately that in order to get the harmonic part of the potential from \req{uads}, one has to choose the $\varphi(\zeta) = \lambda \zeta^2$, and one obtains in this case
\be \lb{uads2}
U_{AdS}(\zeta) = \lambda^2\, \zeta^2 + 2 \lambda (J-1) .
\ee
We can thus extend the superconformal approach to mesons with quark spin $\mathcal{S}=1$ and $J=L+1$ by adding the term $2 \lambda\,\mathcal{S}$ to \req{ulf}, to recover the result \req{uads2} both for mesons with $\mathcal{S}=0$ and $\mathcal{S}=1, \, J=L+\mathcal{S}$.
Therefore our final result for the LF potential, valid for mesons with $J=L+\mathcal{S}$ is:
\be \lb{ulf2}
U(\zeta)= \lambda^2\,\zeta^2 + 2 \lambda (L+\mathcal{S} -1) ,
\ee
and the resulting meson spectrum is the one given by \req{spmes}.

The implementation of the superconformal algebra implies~\cite{deAlfaro:1976vlx, Fubini:1984hf} that besides the Hamiltonian for the bosonic wave function, there is also one for a fermionic one, which describes the supersymmetric fermion of the boson described by \req{ulf}.  Its potential is:
\be \lb{ulffer}
U(\zeta)= \lambda^2 \,\zeta^2 + 2  \lambda (L+1) ,
\ee
and leads to the eigenvalues~~\cite{deTeramond:2014asa,Dosch:2015nwa}
\be \lb{M2lap}
M^2 = 4 \lambda (n + L + 1).
\ee
Consequently  a $q \bar q$ meson with angular momentum $L_M$ has the same mass as a baryon with angular momentum $L_B=L_M+1$ between its quark and diquark cluster components.   This relation has been tested and is very well satisfied for many spectra of light and even heavy hadrons \cite{Dosch:2015nwa,Brodsky:2016yod,Dosch:2016zdv,Nielsen:2018uyn}. The superconformal baryon potential \req{ulffer} can also be obtained from an  AdS  action for fermion fields if a Yukawa-like term $\bar \Psi z \Psi$ is added to the Lagrangian.  This modification had been introduced earlier for purely phenomenological reasons~\cite{Kirsch:2006he, Abidin:2009hr}. The Hamiltonian \req{ulffer} in this case applies to the positively aligned chirality component $\psi^{+}$ of the baryon. There is also the negative chirality component  $\psi^{-}$ of the baryon. The corresponding LF potentials are~\cite{deTeramond:2014asa}
\be U_+(\zeta)= \lambda^2 \,\zeta^2 + 2  \lambda (L+1) \,,
\ee
namely Eq.~\req{ulffer} for the positive component, and
\be
U_-(\zeta)= \lambda^2 \,\zeta^2 + 2  \lambda L \,,
\ee
for the negative component.  The LF potential, together with the term $2 \lambda\, \mathcal{S}$ introduced above, leads to the baryon spectrum given in \req{spbar}.

Finally,  there is also a bosonic superpartner of the negative chirality component $\psi^{-}$ of the baryon, which is interpreted as a tetraquark~\cite{Brodsky:2016yod,Nielsen:2018uyn,Zou:2018eam}. Its mass spectrum is given in \req{sptet}.  The meson, positive and negative  chiral baryon states $\psi^{\pm}$,  and the tetraquark form a 4-plet supermultiplet.

We can generalize the results (\ref{spmes},\ref{spbar},\ref{sptet}) obtained in supersymmetric LFHQCD by going back to normal LFHQCD. Once the modification is fixed, we can use the specific AdS  result \req{uads2} and obtain for the meson spectrum Eq.~\req{spmes2}.


\end{document}